\documentclass[aps,prl,twocolumn,superscriptaddress,amsmath,nofootinbib,amssymb,10pt]{revtex4}
\usepackage{graphicx}
\usepackage{ulem}
\usepackage{soul}      
\usepackage{color}
\usepackage[colorlinks, linkcolor=blue]{hyperref}
\usepackage{verbatim}
\usepackage{esint}
\usepackage{soul}
\renewcommand{\vec}[1]{\boldsymbol{#1}}
\def \rar{{\rightarrow}}
\def \uar{{\uparrow}}

\def \k {{\vec k}}

\def \bI{{\vec I}}
\def \bJ{{\vec J}}

\def \ve {\varepsilon}
\def \r {{\vec r}}

\def \q {{\vec q}}

\def \ve {\varepsilon}

\def \beq {\begin{eqnarray}}
\def \eeq {\end{eqnarray}}
\def \tn {\textnormal}

\def \F {{\cal {F}}}

\def \la{\langle}
\def \ra{\rangle}

\def \cb {\color{blue}}
\def \ie {{\it i.e.}}
\begin{document}

\title{Superconductivity, pseudogap, and phase separation in topological flat bands: \\
a quantum Monte Carlo study}

\author{Johannes S. Hofmann}
\author{Erez Berg}
\email{erez.berg@weizmann.ac.il}
\affiliation{Department of Condensed Matter Physics, Weizmann Institute of Science, Rehovot, 76100, Israel.}
\author{Debanjan Chowdhury}
\email{debanjanchowdhury@cornell.edu}
\affiliation{Department of Physics, Cornell University, Ithaca, New York 14853, USA.}

\begin{abstract}
We study a two-dimensional model of an isolated narrow topological band 
at partial filling with local attractive interactions. Numerically exact quantum Monte Carlo calculations show that the ground state is a superconductor with a critical temperature that scales nearly linearly with the interaction strength. We also find a broad pseudogap regime at temperatures above the superconducting phase that exhibits strong pairing fluctuations and a tendency towards electronic phase separation; introducing a small nearest neighbor attraction suppresses superconductivity entirely and results in phase separation. 
We discuss the possible relevance of superconductivity in this unusual regime to the physics of flat band moir\'{e} materials, and as a route to designing higher temperature superconductors.  
\end{abstract}

\maketitle

{\it Introduction.-} What is the highest attainable superconducting temperature $T_c$ in a given system? This decades-old question has become all the more pressing with the recent discovery of superconductivity in two-dimensional materials with moir\'{e} superlattices~\cite{Cao2018,AY19,Efetov19,FW19,PK19}, which offer an unprecedented degree of controllability of the electronic band structure and  density. It is natural to ask what sets $T_c$ in these systems, as a step towards optimizing it further. 

In general, $T_c$ is limited by two different energy scales: the pairing scale associated with Cooper pair formation, and the phase ordering (or phase coherence) scale, set by the superconducting phase stiffness~\cite{emery1995importance}. Optimizing one energy scale often comes at the expense of the other. For example, in the paradigmatic attractive Hubbard model, increasing the strength of the attractive interaction beyond a certain limit decreases the phase ordering temperature; the optimal $T_c$ is achieved when the attractive interaction $U$ and the electronic bandwidth $W$ are comparable, and the maximum attainable $T_c$ is found to be about $0.02~W$~\cite{RS04,MR10}. 

Intriguingly, it has been suggested that in certain cases, superconductivity can survive even in the limit where the active electronic bands become perfectly flat~\cite{Shaginyan1990,Volovik2011,Kopnin2011,Kopnin2011b,Volovik2013}. In this case, as long as the interaction strength is much smaller than the gap between the active narrow band and the other, empty or filled bands, one expects $T_c$ to be proportional to $U$, which is effectively the only energy scale in the problem. The phase stiffness need not vanish even as the bandwidth vanishes, as long as the single-particle states cannot all be tightly localized \cite{Vanderbilt1,Vanderbilt2}, as in the case, for example, for topological bands. Note that in this case, upon projecting the problem to the active flat bands, the recently proven upper bound on the phase stiffness~\cite{Hazra18} in terms of the bandwidth of the isolated band does not apply, unless contributions from the remote bands are also included~ \cite{MRcomment}.
Interestingly, in several moir\'{e} systems where superconductivity is found, the active bands have been argued to have a topological character~\cite{TS1,TS2,TS3,AB1,BJY1,Wang1,WangChern}.

\begin{figure}[t]
	\begin{center}
		\includegraphics[width=0.99\columnwidth]{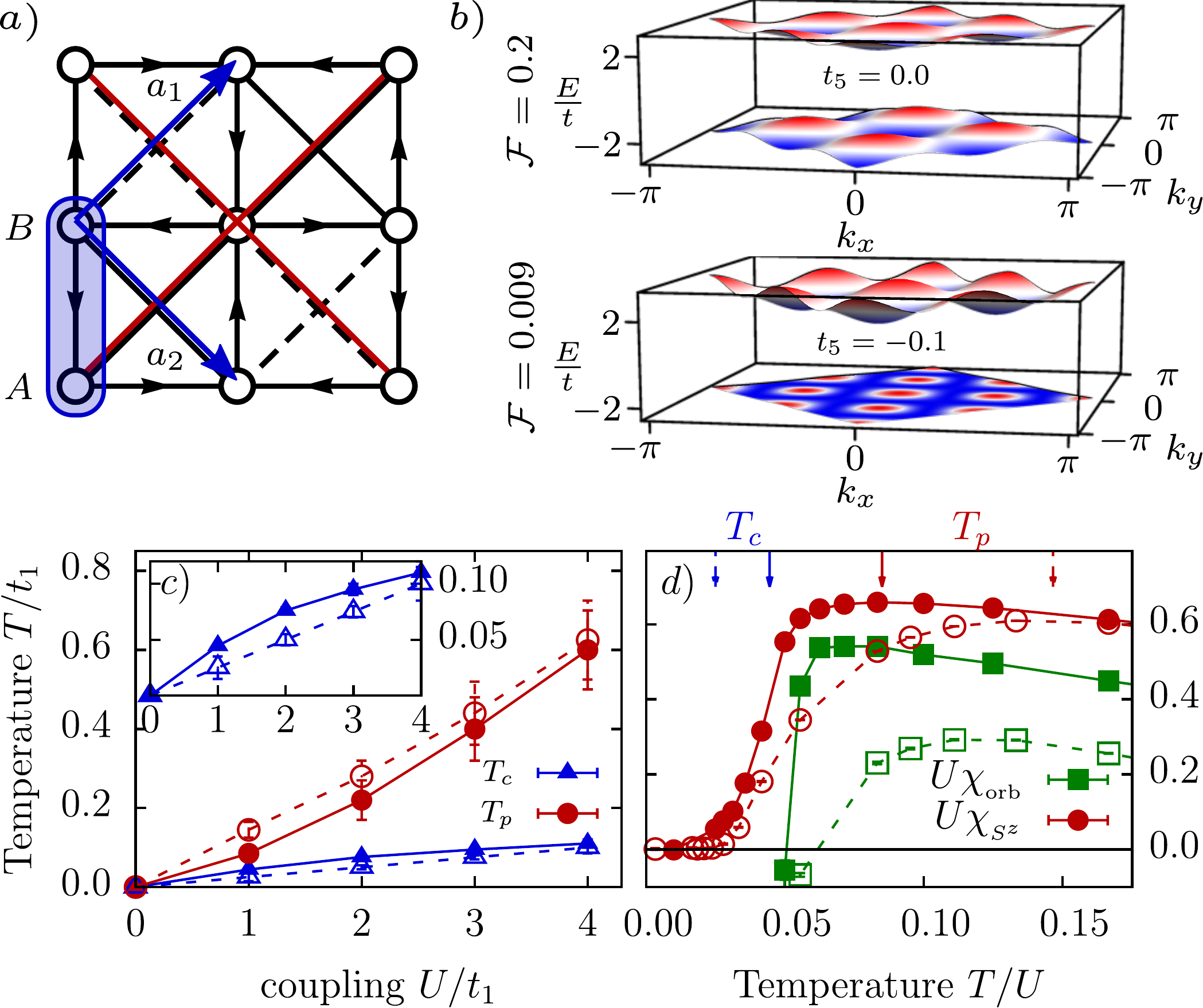}
	\end{center}
	\caption{(a) Lattice model with $\pi$-flux through every plaquette, two orbitals ($A$ and $B$), and first, second, and fifth nearest neighbor hopping of amplitude $t_1$, $t_2$, and $t_5$. 
		(b) Top (bottom) band dispersions for $t_5=0.0$ ($-0.1$) with `flatness ratios' $\mathcal{F} = 0.2$ ($0.009$). The lower bands have Chern numbers $C=+1$ ($-1$) for spin up (spin down) particles. Red (blue) indicates high (low) energy. 
		(c) The superconducting $T_c$ (in blue) and the `pseudogap temperature', $T_p$ (in red, defined through the maximum in the spin susceptibility - see panel d), as a function of $U$. Solid and dashed lines correspond to the band structures shown in b), with $\mathcal{F}=0.2$ and $0.009$, respectively. 
		(d) The orbital and spin magnetic susceptibilities for the dispersive (flat) band with $U=1$ ($U=3$).
	}
	\label{fig:lattice}
\end{figure}

Within Bardeen-Cooper-Schrieffer (BCS) mean-field theory, lower bounds on the phase stiffness in a topological band have been proven~\cite{Torma15,Bernevig19,Torma19,Rossi19}; however, in the limit of a flat band, the problem is inherently strongly coupled and BCS mean-field theory is generally uncontrolled~\cite{BCSremark}. In particular, in this limit all sorts of competing electronic orders may arise (such as charge order and electronic phase separation), and suppress the superconducting $T_c$. While studies of superconductivity in flat bands have been performed~\cite{Scalettar2014,Torma15,Torma16a,Torma16b,Torma17}, superconductivity with $T_c\propto U$ has not been rigorously demonstrated in a solvable model. In addition, the nature of the normal (non-superconducting) state out of which such a superconductor may arise has not been clarified.

In order to address these fundamental questions, we study a sign-problem free lattice electronic-model (Fig.~\ref{fig:lattice}a) with partially filled, narrow-bandwidth Chern bands (Fig.~\ref{fig:lattice}b) with Chern numbers $C=\pm 1$ in the regime of strong attractive interactions using the numerically exact, unbiased determinant quantum Monte-Carlo (DQMC) method \cite{blankenbecler81,alf_v1}. 
It has recently been pointed out that the isolated flat bands in magic-angle twisted bilayer graphene can be decomposed into a total of four $C = 1$ and four $C = -1$ bands~\cite{AV19}.  
Moreover, in a particular solvable limit~\cite{GTAV19}, these Chern bands are tied to a particular sublattice polarization. While the model we study here hosts only two flat $C=\pm1$ bands and does not directly describe the low-energy physics of any particular material, our study serves as a proof-of-principle for addressing many of the questions raised above, in addition to paving the way for building more realistic models for future studies.

We summarize our main findings as follows: (i) For purely on-site interactions, the ground state is an s-wave superconductor, and in the limit where the electronic bandwidth $W$ is much smaller than $U$, there is a broad regime of parameters where $T_c \propto U$ (Fig.~\ref{fig:lattice}c). (ii) Above $T_c$, a broad ``pseudogap'' regime is found, characterized by the opening of a spin-gap (Fig.~\ref{fig:lattice}c,d) and a gap to single-electron excitations (Fig.~\ref{fig:single_particle}) without long-range superconductivity. This regime is characterized by two competing tendencies towards superconductivity and towards electronic phase separation (the latter is signalled by an enhanced electronic compressibility), as a consequence of an approximate emergent SU(2) symmetry at low energies~\cite{Torma16b}. (iii) Adding a small nearest neighbor attraction breaks the SU(2) symmetry and drives an instability to phase separation, thereby destroying SC.

{\it Model.-} We consider the Hamiltonian, $H = H_{\tn{kin}} + H_{\tn{int}}$, defined on a 2D square lattice:
\beq
H_{\tn{kin}} &=& \bigg[-t_1\sum_{\la i,j\ra,\sigma} e^{i\phi^\sigma_{ij}} c^\dag_{i,\sigma} c^{\phantom{\dag}}_{j,\sigma}  - t_2 \sum_{\la i,j\ra_2,\sigma} s_{\la i,j\ra_2} c^\dag_{i,\sigma} c^{\phantom{\dag}}_{j,\sigma} \nonumber \\
&& - t_5 \sum_{\la i,j\ra_5,\sigma} c^\dag_{i,\sigma} c^{\phantom{\dag}}_{j,\sigma} + \tn{h.c.}\bigg] -\mu\sum_i n_i  \\
H_{\tn{int}} &=& -\frac{U}{2}\sum_i (n_i - 1)^2.
\eeq
Here, $c^\dagger_{i,\sigma}$ ($c_{i,\sigma}$) are fermion creation (annihilation) operators, $n_i=\sum_\sigma c^\dagger_{i\sigma}c_{i\sigma}$ is the local density, and $t_1,~t_2,~t_5$ denote the first, second and fifth nearest neighbor hopping parameters (see Fig.~\ref{fig:lattice}a), respectively. The single particle part of the Hamiltonian is a generalization of the model introduced in Ref.~\cite{Neupert11}, designed to give flat Chern bands with Chern numbers $C=\pm 1$. The arrows along the $t_1$ bonds in Fig.~\ref{fig:lattice}a mark the direction associated with $\phi_{ij}^\uar{} = + \frac{\pi}{4}$, and the solid (dashed) second neighbor bonds (whose strength is $t_2$) have a positive (negative) sign $s_{\la\la i,j\ra\ra}$. The red bonds denote $t_5$. The density can be tuned by the chemical potential, $\mu$. The phases satisfy $\phi_{ij}^\sigma = -\phi_{ij}^{-\sigma}$, such that time-reversal symmetry is preserved and $\phi_{ij}^\uar{} = \pm \frac{\pi}{4}$ such that each plaquette encloses $\pi$-flux. $U>0$ is the strength of a local attractive interaction.

It is convenient to define the vectors ${\vec a}_1\equiv(1,1)$ and ${\vec a}_2\equiv(1,-1)$; $\k$ then denotes momenta in the Brillouin zone dual to the lattice spanned by ${\vec a}_1,~{\vec a}_2$  (see Fig.~{\cb{1a}}). $H_{\tn{kin}}$ can be written as,
\beq
H_{\tn{kin}} = \sum_\k \Psi_\k^\dagger \hat{H}_\k \Psi_\k,~~\hat{H}_\k = B_{0,\k}\vec{1} + \vec{B}_\k\cdot\vec\tau,
\label{ham}
\eeq
where $\Psi^\dagger_\k = (c_{\k,A}^\dagger~~c_{\k,B}^\dagger)$ and $\vec\tau\equiv(\tau_x,\tau_y,\tau_z)$ are the Pauli-matrices that act on the sublattice index ($A, B$). 

This leads to two bands, $\ve_\k = B_{0,\k} \pm |\vec{B}_\k|$~\cite{supp}. For the remainder of this study, we fix our hopping parameters $t_1=1$, $t_2=1/\sqrt{2}$ and measure all quantities in units of $t_1$. For $t_5=0$, the gap between the upper and lower band is $\Delta_{\tn{gap}}=4$ and the bandwidth of the lower band is $W=0.828$ (Fig.~\ref{fig:lattice}b); the `flatness-ratio', $\F \equiv W/\Delta_{\tn{gap}}= 0.2$. 
We can tune the bandwidth, and thereby $\F$, of the lower Chern band by varying $t_5$. The flatness-ratio is minimized by $t_5=\frac{1-\sqrt{2}}{4}$ where the bandwidth for the lower band, $W\approx0.035$, while the gap remains at $\Delta_{\tn{gap}}=4$, such that $\mathcal{F} = 0.009$. For most of our study, we focus on the following parameter values: (a) $\F=0.2$, and, (b) $\F=0.009$ for a range of values between $U=1 - 4$, and  the case of quarter-filling ($\nu = 1/4$), corresponding to a half-filled (lower) Chern band.

\begin{figure}
	\begin{center}
		\includegraphics[width=0.99\columnwidth]{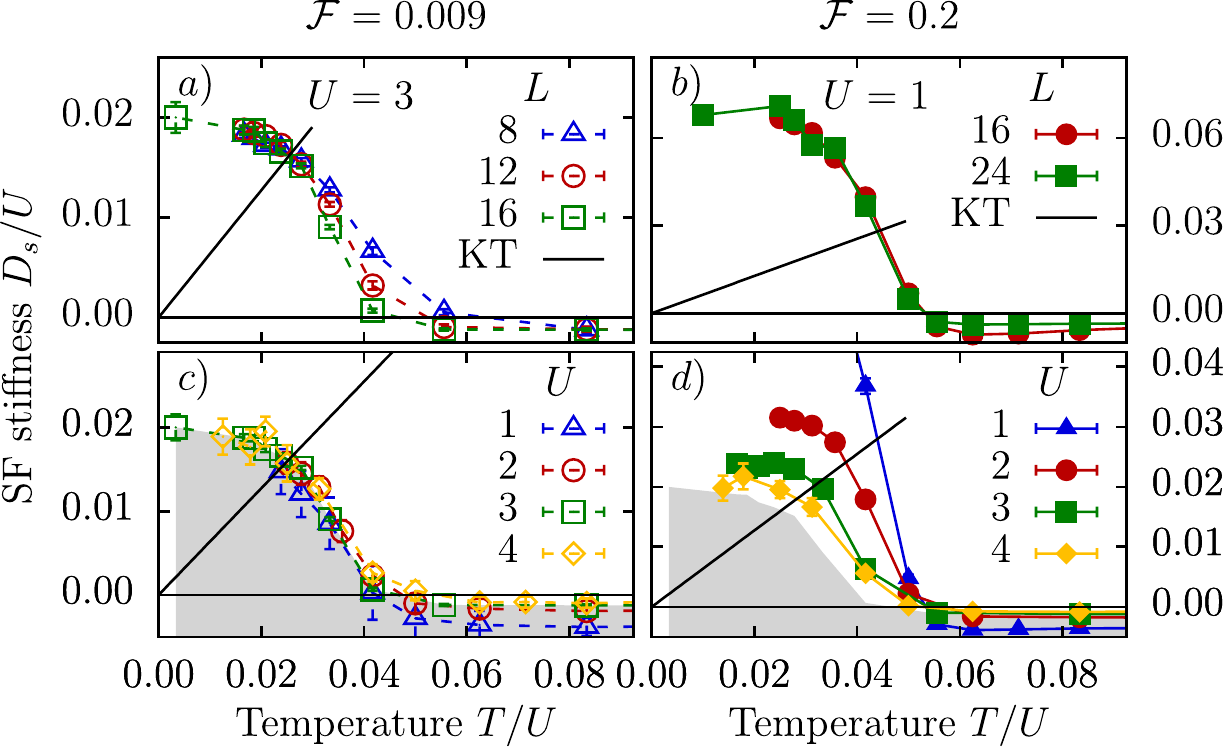}
	\end{center}
	\caption{The superfluid stiffness $D_s(T)$ [Eq.~\eqref{Rhos}] for the (a) flat-band ($\mathcal{F}=0.009$, $U=3$), and (b) dispersive ($\mathcal{F}=0.2$, $U=1$) cases, and different system sizes. 
		The black line denotes the universal BKT jump, $D_s = 2 T/\pi$. 
		(c) and (d): $D_s(T)$ for various coupling strength $U$ on the largest simulated lattice. 
		In (c), $D_s(T)/U$ curves for different $U$'s collapse onto each other when plotted vs. $T/U$, confirming that $U$ is effectively the only energy scale in the flat band case. The shaded area marks the collapsed function in both (c) and (d) to guide the eye for comparison.
	}
	\label{fig:superconductivity}
\end{figure}

\begin{figure}
	\begin{center}
		\includegraphics[width=0.99\columnwidth]{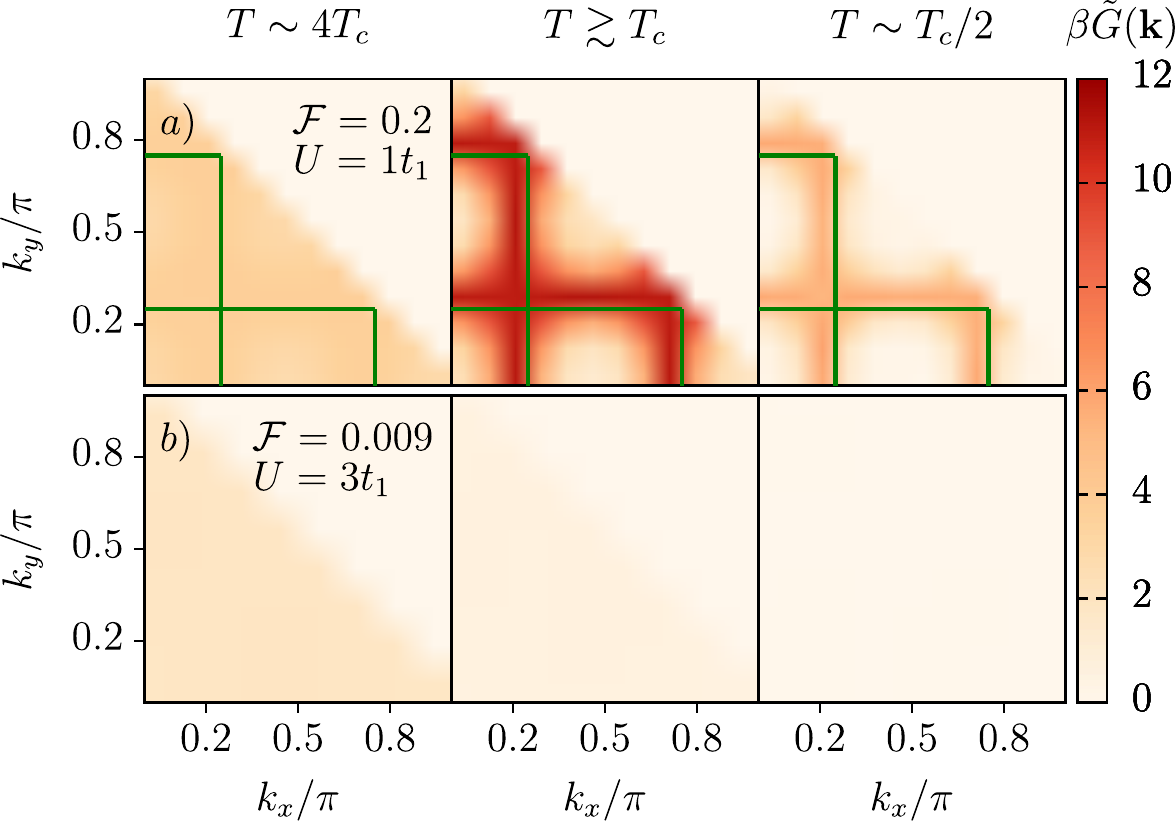}
	\end{center}
	\caption{The quantity $\beta \tilde{G}(\k)$, that serves as a proxy for the spectral function near the Fermi energy [see Eq.~(\ref{eq:G}) and the following discussion], as a function of $\k$. The green lines denote the Fermi surface in the non-interacting case. {In (a), $U=1$ and the temperatures are $T=0.25$, $0.06$, $0.03$; in (b), $U=3$  and $T=0.4$, $0.1$, $0.05$.} 
	}
	\label{fig:single_particle}
	
\end{figure}

{\it Superconductivity.-} In order to diagnose the possible onset of SC, we compute the phase stiffness $D_s$.
We evaluate the paramagnetic current-current correlation function, $\Lambda_{xx}(\q,i\omega_m=0)$, at zero external Matsubara frequency and use the relation~\cite{SFcriteria,RS04},
\beq
\label{Rhos}
D_s = \frac{1}{4}\big[- K_x - \Lambda_{xx}(\q=0) \big].
\eeq
Here, $K_x = \left\langle \left[ \partial^2 H/\partial A_x^2\right]_{A_x=0}\right\rangle$ is the diamagnetic current contribution, where $A_x$ is a vector potential in the $x$ direction, introduced via minimal coupling, and the $1/4$ prefactor is due to charge-2 Cooper pairs~\cite{supp}.
We plot $D_s(T)$ as a function of temperature in Fig.~\ref{fig:superconductivity}a, b. The chemical potential $\mu(T)$ is tuned such that $\nu=1/4$. In 2D, $T_c$ can be determined from the Berezinskii-Kosterlitz-Thouless (BKT) condition $T_c=\pi D_s^-/2$, where $D_s^-\equiv D_s(T\rar T_c^-)$. 
The black solid line denotes the curve $D_s = 2T/\pi$, the intersection of which with $D_s(T)$ gives $T_c$. The $T_c$ values extracted from $D_s(T)$ are consistent with an independent analysis of the superconducting correlation length $\xi_{\tn{SC}}/L$~\cite{supp}. 

The slightly negative $D_s$ values found at high temperatures are associated with Trotter errors, and we have checked that they decrease in magnitude towards zero upon decreasing the imaginary time step $\Delta \tau$. 
We have also confirmed the absence of a few possible competing orders such as a charge density wave, a bond density wave or magnetic states~\cite{supp}.

The BKT transition temperature as a function of $U$ is shown in the inset of Fig.~\ref{fig:lattice}c for the two band structures with $\mathcal{F}=0.2$, $0.009$. Most strikingly, for the narrow band ($\mathcal{F}=0.009$), $T_c$ depends almost perfectly linearly on $U$: $T_c \approx 0.025 U$. In the case of the more dispersive band, $\mathcal{F}=0.2$, $T_c$ is higher than for the narrower band, and has a downward curvature. As $U$ increases, the $T_c$'s of the two band structures approach each other. This behavior can be understood in terms of two contributions to the phase stiffness: (i) a geometric contribution originating from the finite extent of the wave functions spanning the topological bands, that does not vanish even in the $W\rightarrow0$ limit, and (ii) the conventional contribution originating from the single-particle kinetic energy.

The dependence of $T_c$ on $U$ is hence markedly different both from the conventional weak-coupling BCS behavior, $T_c\sim W e^{-\frac{W}{U}}$, 
and from the strong coupling behavior found in the attractive Hubbard model, $T_c\sim W^2/U$. To shed more light into the origin of this behavior, we present in Fig.~\ref{fig:superconductivity} c, d  scaling plots of $D_s/U$ as a function of $T/U$ for different values of $U$. For the narrower band (panel c), the different curves collapse on top of each other. This can be understood by considering the limit $W \ll T \ll \Delta_\tn{gap}$. Since the upper band can effectively be projected out in this regime, the superfluid density must be of the form $D_s = U f(T/U, \nu)$, where $f$ is a scaling function that depends only on the Bloch wavefunctions of the lower band~\cite{supp}. Fixing $\nu$ gives a scaling collapse of the form observed in Fig.~\ref{fig:superconductivity}c. For the more dispersive case (panel d) the curves do not collapse. As $U$ increases, however, the $\mathcal{F}=0.2$ curves converge towards the shaded form, which is the scaling function for $\mathcal{F}=0.009$. 

{\it Normal-state properties.-} Let us now examine the properties of the normal (non-superconducting) state for $T>T_c$. In the limit where the bare band is very narrow, the key question is whether the normal state should be understood in terms of coherent quasi-particle excitations whose bandwidth is set by the interaction strength, or as an incoherent liquid of Cooper pairs~\cite{Tovmasyan2018,Wang2019}. As described below, our findings are consistent with the latter scenario: as $\mathcal{F}$ decreases, a broad ``pseudogap'' regime appears above $T_c$, characterized by the opening of a gap for spin and single-particle excitations. The pseudogap regime further displays strong superconducting fluctuations and a tendency towards phase separation.

In order to probe the single electron spectral function,  $A(\k,\omega)=-\pi^{-1} \textrm{Im}\, G(\k,\omega)$, we recall that the imaginary time Green's function, $G(\k,\tau)=\sum_{\alpha=A,B} \la c^{\phantom{\dag}}_{\alpha\k}(\tau)c^\dag_{\alpha\k}(0) \ra$, for $0<\tau<\beta$ has the following property~\cite{Trivedi},
\beq
G(\k,\tau) = \int_{-\infty}^{\infty} d\omega \frac{e^{-\omega(\tau-\beta/2)}}{2\cosh(\beta\omega/2)} A(\k,\omega).
\label{eq:G}
\eeq
Thus, $\widetilde{G}(\k)\equiv G(\k,\tau=\beta/2)$ is the integrated spectral weight around the Fermi level over a width of $\sim T$. In particular, $\lim_{T\rightarrow 0}\beta\widetilde{G}(\k)=\pi A(\k,\omega=0)$. Fig.~\ref{fig:single_particle}a,b shows the evolution of $\widetilde{G}(\k)$ as a function of decreasing temperature from $T\sim 4T_c$ down to $T\sim T_c/2$ for two parameter sets, $(\mathcal{F}, U) = (0.2,1 )$ and $(0.009,3 )$. 

For the more dispersive band ($\mathcal{F}=0.2$, Fig.~\ref{fig:single_particle}a), $\widetilde{G}(\k)$ is peaked near the original non-interacting Fermi surface (but is significantly broadened). Moreover, even in the SC state at $T\sim T_c/2$, when the Fermi surface develops a SC gap, the remnant of the gapped Boguliubov spectrum continues to remain visible near the original Fermi surface. On the other hand, for the flatter band ($\mathcal{F}\sim 0.009$) at stronger-coupling, $\widetilde{G}(\k)$ is completely featureless across $T_c$, showing no sign of coherently propagating quasi-particles nor a well defined Fermi surface. Hence, the superconductivity here cannot be understood as a Fermi surface instability. Instead, we will show in the remainder that it emerges from an incoherent liquid of preformed pairs.

The normal state is further characterized by its charge, magnetic (Zeeman and orbital), and pairing susceptibilities, defined as:
\beq
\chi_{\hat{O}}&=&L^{-2} \int_0^\beta d\tau \la \hat{O}(\tau)\hat{O}(\tau=0) \ra \\
\chi_{\tn{orb}}&=&\lim_{q\rightarrow 0} q^{-2} \big[\Lambda_{xx}^{\tn{t}}(q)-\Lambda_{xx}^{\tn{l}}(q)\big],
\eeq
with $\hat{O}$ being the total $z$-component of the spin ($S^z = \sum_{j} c_j^\dagger \sigma^z c_j$), charge ($N = \sum_{j} (n_j-\nu)$), and s-wave pairing ($\Delta = \sum_{j} c_{j,\uparrow} c_{j,\downarrow} + \tn{h.c.}$), respectively. For the orbital magnetic susceptibility, we use the notation $\Lambda_{xx}^{\tn{t}}(q)=\Lambda_{xx}(q_x=0,q_y=q)$ and $\Lambda_{xx}^{\tn{l}}(q)=\Lambda_{xx}(q_x=q,q_y=0)$ for the transverse and longitudinal components~\cite{supp}. 

The spin and orbital magnetic susceptibilities are presented in Fig.~\ref{fig:lattice}d. The spin-susceptibility, $\chi_{S^z}$, shows a clear suppression below a characteristic temperature scale, indicating the onset of a spin gap. We define the ``pseudogap temperature'' $T_p$ as the location of the maximum of $\chi_{S^z}(T)$, is shown in Fig.~\ref{fig:lattice}c as a function of $U$, and is found to be substantially above $T_c$ at strong coupling.  
$\chi_\tn{orb}$ is positive (paramagnetic) at high temperature, but drops sharply and becomes large and negative (diamagnetic) at a temperature above $T_c$. The sign change in $\chi_{\text{orb}}$ occurs at $T\approx 0.05 U$ (Fig.~\ref{fig:lattice}d). This behavior can be understood as the consequence of the onset of pairing fluctuations, which give a diamagnetic contribution to the orbital susceptibility.

\begin{figure}
	\begin{center}
		\includegraphics[width=0.99\columnwidth]{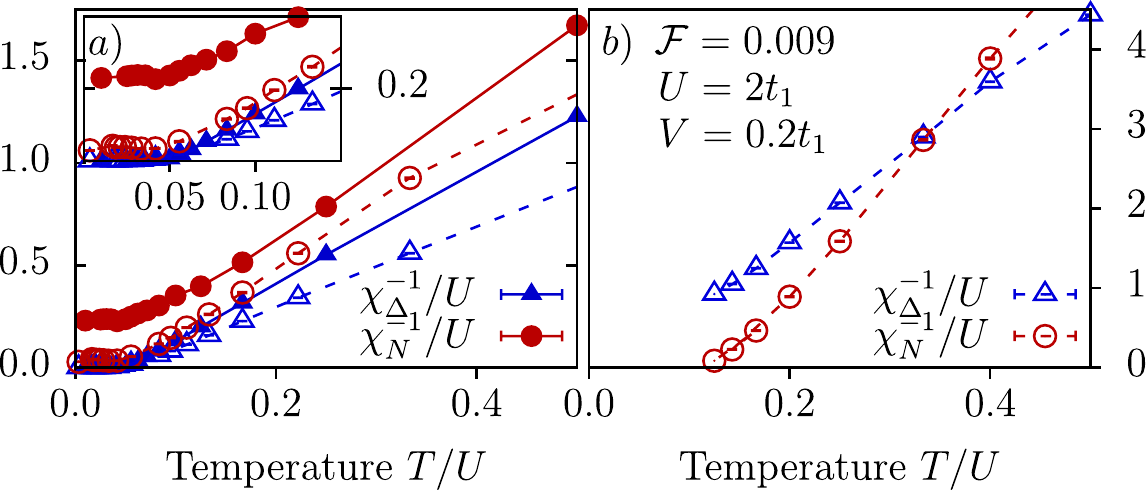}
	\end{center}
	\caption{
		(a) The reciprocal pairing and charge susceptibilities are shown for $\mathcal{F}=0.2$, $U=1t_1$ (solid) and $\mathcal{F}=0.009$, $U=3t_1$ (dashed). (b) same as (a) but with an additional nearest-neighbor interaction $H_{\tn{nn}} = -V\sum_{\la i,j\ra} (n_i-1) (n_j-1)$, $V=0.2t_1$.
	}
	\label{fig:susceptibilities}
	
\end{figure}

Finally, we present the reciprocal pairing and charge susceptibilities, $\chi_{\Delta}^{-1}$, $\chi_N^{-1}$, in Fig.~\ref{fig:susceptibilities}a. For a broad range in temperature below $\Delta_{\tn{gap}}$ and above $T_c$, the pairing susceptibility appears to follow a Curie-Weiss law $\chi_\Delta\sim (T-\Theta)^{-1}$. 
Strikingly, the charge susceptibility $\chi_N$ is also strongly enhanced in the same temperature regime.
This signals a tendency towards phase separation, driven by the same attractive interaction that is responsible for superconductivity. Phase separation is ultimately preempted by superconductivity, however, and $\chi_N$ saturates below $T_c$. The enhancement of $\chi_N$ with decreasing temperature is particularly strong for the narrower band ($\mathcal{F}=0.009$).
This can be understood as a consequence of an emergent SU(2) symmetry 
in the limit $\mathcal{F}\rightarrow0$ and $U/\Delta_{\tn{gap}}\rightarrow0$~\cite{Torma16b,supp}. 
In this limit, the BCS wave function is an exact ground state. The SU(2) symmetry  relates the superconducting susceptibility to the charge susceptibility; hence, $\chi_\Delta =  \chi_N$, and both diverge in the limit $T\rightarrow 0$. 

In our system,
the SU(2) symmetry is weakly broken, both due to the finite $U/\Delta_{\mathrm{gap}}$ and the non-zero bandwidth~\cite{supp}. This breaking of the symmetry tilts the balance in favor of superconductivity, rendering $T_c$ finite and saturating $\chi_N$. 
Interestingly, in the case of the more dispersive band, $\chi_N$ continues to be temperature dependent even for $T<W=0.828t_1$. This is reminiscent of the behavior observed in the repulsive Hubbard model at intermediate temperatures~\cite{Bakr,Huang2018}. 

The close competition between superconductivity and phase separation suggests that the superconducting state is fragile to small perturbations. To demonstrate this fragility, we studied the effect of adding nearest-neighbor interactions to our original Hamiltonian, $H_{\tn{int}}\rightarrow H_{\tn{int}} - V\sum_{\la i,j\ra}(n_i-1)(n_j-1)$. Fig.~\ref{fig:susceptibilities}b shows $\chi^{-1}_\Delta(T)$, $\chi^{-1}_N(T)$ upon switching on $V=0.1U=0.2t_1$. The nearest-neighbor interaction drives a finite-temperature instability towards phase separation, signalled by $\chi^{-1}_N\rightarrow 0$, that preempts the superconducting transition. This fragility of the superconducting state is a consequence of the approximate SU(2) symmetry; the nearest-neighbor attraction breaks the symmetry and favors phase separation~\cite{supp}. Note that this is a strong coupling effect, not attainable within a BCS treatment of the problem.

{\it Discussion \& Outlook.-} We have demonstrated explicitly that superconductivity is possible in the limit of a nearly flat bare band in the presence of local attractive interactions. In this strong coupling regime, the interaction strength is the dominant energy scale in the problem; consequently, $T_c\propto U$. Moreover, the superconducting state emerges from a pseudogap regime, where single-particle and spin excitations are gapped, and superconducting as well as particle number fluctuations are strongly enhanced. As a result of an approximate SU(2) symmetry, a minimal low-energy response can be captured in the intermediate temperature regime in terms of the thermal fluctuations of a non-linear sigma model (NLSM) for a multi-component order parameter. However, as a result of the weak SU(2) symmetry breaking, the order parameter manifold is not perfectly symmetric and the NLSM needs to be supplemented with a slight easy-plane anisotropy, which may favor either SC or phase separation~\cite{supp}.

Clearly, an essential ingredient for superconductivity in the flat band regime is the geometric character of the band; it is crucial that the wavefunctions spanning the band are not completely localizable~\cite{Torma15,Bernevig19}. 
An interesting open question, worthy of further investigations, is to what extent is band topology essential for superconductivity in this regime.

Finally, we speculate about the relevance of the physics discussed here to superconductivity in two-dimensional moir\'{e} materials. In these systems, superconductivity is indeed found in extremely narrow, topologically non-trivial bands. It would be interesting to look for a pseudogap regime above the superconducting $T_c$, characterized by strong pairing fluctuations and an enhanced electronic compressibility. Incidentally, indirect signatures of a possible pseudogap above $T_c$ have been reported in twisted bilayer graphene~\cite{SMBLG}.

{\it Acknowledgements.-} The authors thank Fakher Assaad, Francesco Parisen Toldin, Tobias Holder, and Mohit Randeria for stimulating discussions. DC is supported by faculty startup funds at Cornell University. DC also acknowledges hospitality of the Weizmann Institute of Science and the Max-Planck Institute for the Physics of Complex Systems. EB and JH were supported by the European Research Council (ERC) under grant HQMAT (grant no. 817799), and by the US-Israel Binational Science Foundation (BSF).
The authors gratefully acknowledge the Gauss Centre for Supercomputing e.V. (www.gauss-centre.eu) for funding this project by providing computing time on the GCS Supercomputer SuperMUC at Leibniz Supercomputing Centre (www.lrz.de) under the project number pr53ju. This work was supported by a research grant from Irving and Cherna Moskowitz.

\bibliographystyle{apsrev4-1_custom}
\bibliography{refs}

\begin{thebibliography}{47}
\bibitem [{MRc()}]{MRcomment}%
  \BibitemOpen
  \href@noop {} {}\bibinfo {note} {The interaction term, when projected to the
  isolated topological bands, becomes non-trivial and contributes to the
  electron current. The stiffness can then no longer be bounded by the kinetic
  energy terms alone.}\BibitemShut {Stop}%
\bibitem [{BCS()}]{BCSremark}%
  \BibitemOpen
  \href@noop {} {}\bibinfo {note} {If the active band is precisely flat and
  particle-hole symmetric, then the BCS wavefunction is exact~\cite{Torma15},
  due to an SU(2) symmetry that relates the charge and pairing operators. Under
  these conditions, $T_c=0$. {An approximate SU(2) symmetry may also emerge at
  low energies under certain conditions~\cite{Torma16b}}, as we discuss in
  detail below.}\BibitemShut {Stop}%
\end{thebibliography}
\clearpage

\renewcommand{\thefigure}{S\arabic{figure}}
\renewcommand{\figurename}{Supplemental Figure}
\setcounter{figure}{0}
\begin{widetext}
\appendix
\section{SUPPLEMENTARY INFORMATION}
\subsection{Band dispersion}
\label{bands}
For the sake of completeness, we specify here the momentum dependencies of the $B^\mu_\k$ introduced in Eq.~\ref{ham} in the main text,
\beq
B_{\k}^x + iB_{\k}^y &=& -2t_1e^{-i\pi/4}e^{- i k_y}\cos(k_y) 
- 2t_1e^{i\pi/4}e^{- i k_y}\cos(k_x),\\
B_{\k}^z &=& -2t_2[\cos(k_x+k_y) - \cos(k_x-k_y)] \\
B_{\k}^0&=& -2t_5[\cos(2(k_x+k_y)) + \cos(2(k_x-k_y))].
\eeq

\subsection{Equal time $s$-wave pair correlation function}
\label{sec:CorrLength}

In addition to the superfluid stiffness, we also diagnose the superconducting phase transition by calculating the equal-time correlation function,
\beq
S_\Delta(\q)=\frac{2}{L^2}\sum_{\r,\r'} e^{-i(\r-\r')\cdot\q} \la \Delta^\dag(\r)\Delta^{\vphantom{\dag}}(\r') \ra,\label{CorrFct}
\eeq
where $\Delta(\r)=c_{\r,\uparrow}c_{\r,\downarrow}$ is an on-site, spin-singlet SC order parameter.
We observe a `Bragg-like' peak for $S_\Delta(\q)$ at the wave vector $\q=0$ (see first column in Fig. \ref{fig:CompOrder}) as a hint of onset of long-range (amplitude) order with uniform SC in the $s-$wave spin-singlet channel. 

\begin{figure}
	\begin{center}
		\includegraphics[width=0.7\columnwidth]{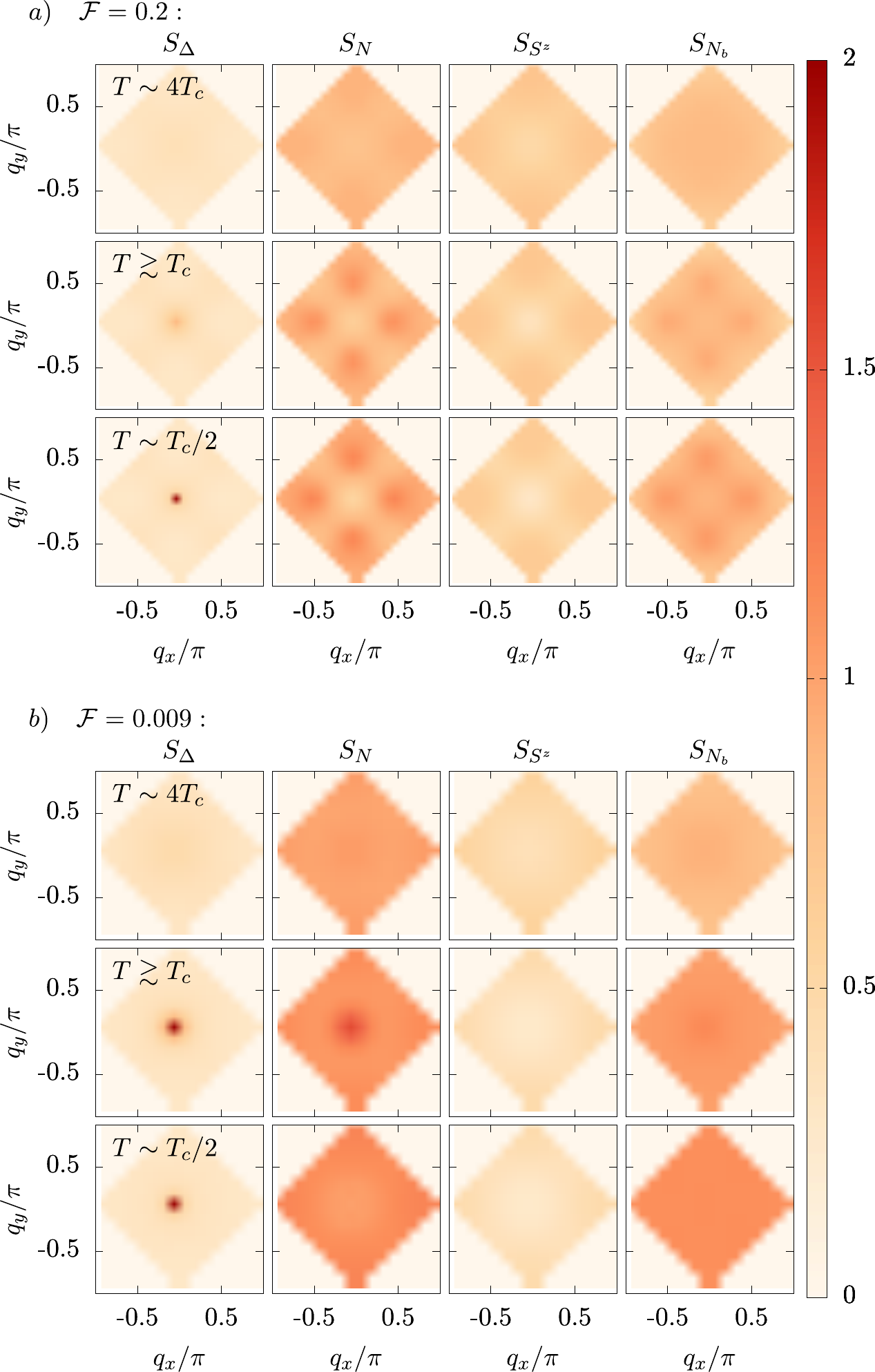}
	\end{center}
	\caption{Momentum resolved equal-time correlation functions of $s$-wave pairing ($\Delta$), charge ($N$), spin ($S^z$), and bond-density ($N_b$) operators at various temperatures for (a) the dispersive band at coupling strength $U=1$ ($L=24$) and (b) the flat band regime with $U=3$ ($L=16$). The $\q=(0,0)$ Bragg peak in the pairing correlation function signals an increased superconducting correlation length. {The broad peak at intermediate temperatures in the density correlation functions of the flat band is a consequence of a tendency towards phase separation, that is ultimately preempted by superconductivity at low temperatures.} The absence of additional sharp peaks
		in all other channels indicates that there is no other competing order. 
	}
	\label{fig:CompOrder}
	
\end{figure}
We extract the correlation length, $\xi_{\tn{SC}}$, from the momentum dependence of $S_\Delta(\q)$ at small wavevectors \cite{Toldin15},
\begin{equation}
\xi_{\tn{SC}}=\frac{1}{2\sin(\pi / L)}\sqrt{\frac{S_\Delta(\q=(0,0))}{S_\Delta(\q=(2\pi/L,0))}-1}.\label{CorrLength}
\end{equation}

Let us begin the analysis by recalling that in BKT transition, the critical temperature $T_c$ separates a disordered phase at high temperatures, which is characterized by exponentially decaying correlation functions and thus by finite $\xi_{\tn{SC}}$, from an algebraic phase at low temperatures, which is characterized by power-law correlations with divergent $\xi_{\tn{SC}}$. From a renormalization-group (RG) perspective, the algebraic phase below $T_c$ is described by a line of scale-invariant fixed points.
Note that phase fluctuations prohibit long-range order at any finite temperature.
By studying the dependence of $\xi_{\tn{SC}}/L$ on $L$ as a function of decreasing temperatures, we can identify $T_c$. In a scale-invariant theory, $\xi_{\tn{SC}}/L$ is lattice-size independent at leading order and may exhibit $\ln(L)$ finite-size scaling corrections. Hence, $T_c$ is marked by the temperature below which the lines in Figs.~\ref{fig:CorrRatio} a, b merge. 

The most noticeable difference between the two panels is a surprisingly clear crossing point  at $T_c/U=0.045\pm0.005$ in Fig.~\ref{fig:CorrRatio} a while the curves for different lattice sizes seem to merge below $T_c=0.025\pm0.007$ in Fig.~\ref{fig:CorrRatio} b. The latter is consistent both with the expectation described above as well as the critical temperature determined in the main text using the universal jump of the superfluid stiffness. The former, however, exhibit much more pronounced finite size effects, even on the comparatively larger lattices, presumably due to the larger kinetic energy of the fermions. Hence, in the case of the more dispersive band, the transition appears to be more `mean-field' like on finite size lattices.

\begin{figure}
	\begin{center}
		\includegraphics[width=0.6\columnwidth]{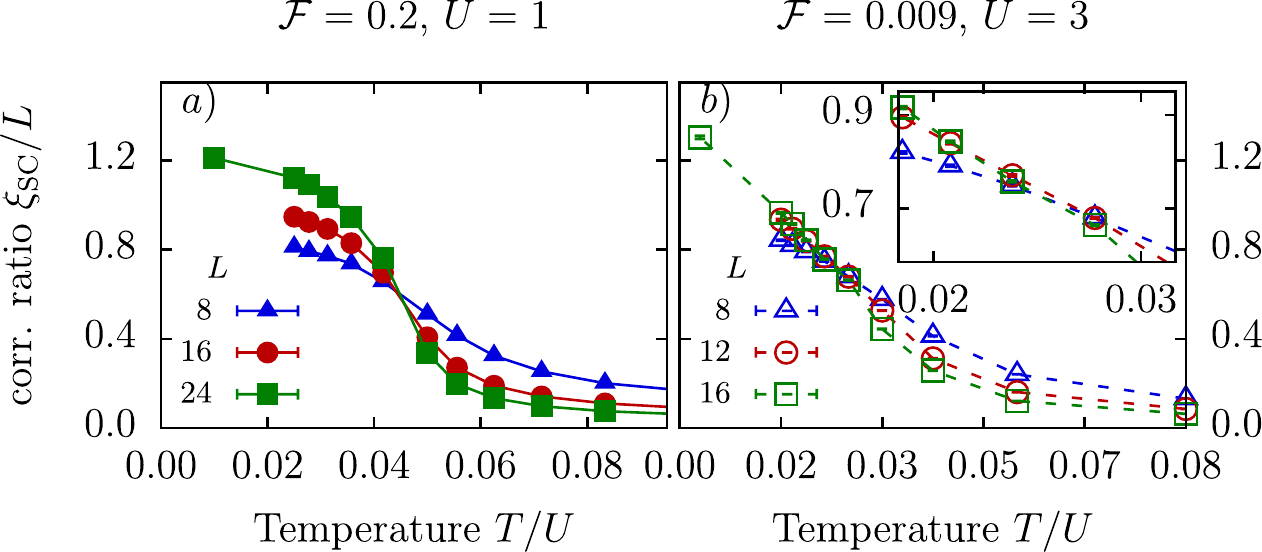}
	\end{center}
	\caption{Correlation ratio $\xi_{\tn{SC}}/L$ as a function of temperature for (a) the dispersive band at coupling strength $U=1$ and (b) the flat band regime with $U=3$. The size-independent crossing/merging point marks the critical temperature $T_c$ of the phase transition. We estimate $T_c/U=0.045\pm0.005$ ($T_c/U=0.025\pm0.007$) in the dispersive (flat) band setup. These values are in broad agreement with the critical temperatures deduced from the SF density (Fig.~\ref{fig:superconductivity} in the main text), which are $T_c/U=0.0445\pm0.0015$ ($T_c/U=0.025\pm0.001$).
	}
	\label{fig:CorrRatio}
	
\end{figure}

\subsection{Electromagnetic response}

\subsubsection{Dia- and paramagnetic current operator}

Each bond is coupled to the electromagnetic field via the usual Peierls' substitution as 
\begin{equation}
c^\dag_{\bI \alpha}c^{\phantom{\dag}}_{\bJ \beta} \rightarrow c^\dag_{\bI \alpha}c^{\phantom{\dag}}_{\bJ \beta} \exp\left[-ie\int_{\r_{\bJ\beta}}^{\r_{\bI \alpha}}{\bf{A}}({\bf{l}})d{\bf{l}}\right]
\end{equation}
where $\bI$, $\bJ$ labels the unit cell of the lattice, $\alpha$, $\beta$ the orbital within the unit cell and $\r_{\bI \alpha}$, $\r_{\bJ\beta}$ are their real-space positions.
We use the long-wavelength approximation and assume that the vector potential is constant for the length of the bond. This yields $\int_{\r_{\bJ\beta}}^{\r_{\bI \alpha}}{\bf{A}}({\bf{l}})d{\bf{l}}={\bf{A}}({\bf{R}}){\bf{d}}$ with $\bf{R}=(\r_{\bI \alpha}+\r_{\bJ\beta})/2$ and $\bf{d}=\r_{\bI \alpha}-\r_{\bJ\beta}$.

Furthermore, we focus on the current response in the $x$ direction such that ${\bf{A}}=A_x\hat{{\bf{e}}}_x$. Hence, the potential does not couple to bonds that are purely in the $y$ direction. Note that both the nearest neighbor bond in the $x$ direction and all next-nearest neighbor bonds couple with $\exp(\pm i e A_x(\bf{R}))$, whereas the fifth-nearest neighbor bond has $\exp(\pm i e 2 A_x(\bf{R}))$. This leads to 10 separate contributions per unit cell $\bI$ for the paramagnetic current operator $J_x$ and the corresponding diamagnetic term $K_x$ (with an implied sum over the spin degree of freedom)

\noindent
\begin{minipage}{0.47\textwidth}
	\begin{subequations}
		\begin{eqnarray}
		J_x^1(\bI)&=& i t e^{i\pi/4} c^\dag_{\bI+{\bf{a}}_2 2}c^{\phantom{\dag}}_{\bI 1} + \tn{h.c.} \\
		J_x^2(\bI)&=& i t_2 c^\dag_{\bI+{\bf{a}}_1 1}c^{\phantom{\dag}}_{\bI 1} + \tn{h.c.} \\
		J_x^3(\bI)&=&-i t_2 c^\dag_{\bI+{\bf{a}}_2 1}c^{\phantom{\dag}}_{\bI 1} + \tn{h.c.} \\
		J_x^4(\bI)&=& i t e^{-i\pi/4} c^\dag_{\bI+{\bf{a}}_1 1}c^{\phantom{\dag}}_{\bI 2} + \tn{h.c.} \\
		J_x^5(\bI)&=&-i t_2 c^\dag_{\bI+{\bf{a}}_1 2}c^{\phantom{\dag}}_{\bI 2} + \tn{h.c.} \\
		J_x^6(\bI)&=& i t_2 c^\dag_{\bI+{\bf{a}}_2 2}c^{\phantom{\dag}}_{\bI 2} + \tn{h.c.} \\
		J_x^7(\bI)&=& i 2 t_3 c^\dag_{\bI+2{\bf{a}}_1 1}c^{\phantom{\dag}}_{\bI 1} + \tn{h.c.} \\
		J_x^8(\bI)&=& i 2 t_3 c^\dag_{\bI+2{\bf{a}}_2 1}c^{\phantom{\dag}}_{\bI 1} + \tn{h.c.} \\
		J_x^9(\bI)&=& i 2 t_3 c^\dag_{\bI+2{\bf{a}}_1 2}c^{\phantom{\dag}}_{\bI 2} + \tn{h.c.} \\
		J_x^{10}(\bI)&=& i 2 t_3 c^\dag_{\bI+2{\bf{a}}_2 2}c^{\phantom{\dag}}_{\bI 2} + \tn{h.c.} \\ \nonumber
		\end{eqnarray}
	\end{subequations}
\end{minipage}
\hfill
\begin{minipage}{0.47\textwidth}
	\begin{subequations}
		\begin{eqnarray}
		K_x^1(\bI)&=& - t e^{i\pi/4} c^\dag_{\bI+{\bf{a}}_2 2}c^{\phantom{\dag}}_{\bI 1} + \tn{h.c.} \\
		K_x^2(\bI)&=& - t_2 c^\dag_{\bI+{\bf{a}}_1 1}c^{\phantom{\dag}}_{\bI 1} + \tn{h.c.} \\
		K_x^3(\bI)&=&   t_2 c^\dag_{\bI+{\bf{a}}_2 1}c^{\phantom{\dag}}_{\bI 1} + \tn{h.c.} \\
		K_x^4(\bI)&=& - t e^{-i\pi/4} c^\dag_{\bI+{\bf{a}}_1 1}c^{\phantom{\dag}}_{\bI 2} + \tn{h.c.} \\
		K_x^5(\bI)&=&   t_2 c^\dag_{\bI+{\bf{a}}_1 2}c^{\phantom{\dag}}_{\bI 2} + \tn{h.c.} \\
		K_x^6(\bI)&=& - t_2 c^\dag_{\bI+{\bf{a}}_2 2}c^{\phantom{\dag}}_{\bI 2} + \tn{h.c.} \\
		K_x^7(\bI)&=& - 4 t_3 c^\dag_{\bI+2{\bf{a}}_1 1}c^{\phantom{\dag}}_{\bI 1} + \tn{h.c.} \\
		K_x^8(\bI)&=& - 4 t_3 c^\dag_{\bI+2{\bf{a}}_2 1}c^{\phantom{\dag}}_{\bI 1} + \tn{h.c.} \\
		K_x^9(\bI)&=& - 4 t_3 c^\dag_{\bI+2{\bf{a}}_1 2}c^{\phantom{\dag}}_{\bI 2} + \tn{h.c.} \\
		K_x^{10}(\bI)&=& - 4 t_3 c^\dag_{\bI+2{\bf{a}}_2 2}c^{\phantom{\dag}}_{\bI 2} + \tn{h.c.} \\ \nonumber
		\end{eqnarray}
	\end{subequations}
\end{minipage}

\noindent with the lattice vectors ${\bf{a}}_1=(1,1)$ and ${\bf{a}}_2=(1,-1)$ (see Fig.~1a).

\subsubsection{Technical remarks on the Fourier transformation}

It is straight forward to calculate the total diamagnetic contribution $K_x=L^{-2}\sum_{\bI,\alpha}K_x^\alpha(\bI)$. To extract the paramagnetic current-current correlation function $\Lambda_{xx}(\q)$, we have to include the spatial resolution within the unit cell. Respecting the two-orbital unit cell, the Fourier transformation is done with respect to the position of the unit cells $\r_\bI \equiv \r_{\bI,A} $ such that we keep the different current terms separate and define the correlation matrix
\begin{equation}
\Lambda_{xx}^{\alpha \beta}(\q)=\sum_{\bI,\bJ} e^{-i(\r_\bI-\r_\bJ)\q} \int_0^\beta d\tau \la J_x^\alpha(\bI,\tau) J_x^\beta(\bJ,\tau=0) \ra\,.
\end{equation}
We can then include the position of the bonds within the unit cell by an additional phase factor $e^{-i\q {\bf{d}}^\alpha}$ such that
\begin{equation}
\Lambda_{xx}(\q)=\sum_{\alpha \beta} e^{-i({\bf{d}}^\alpha-{\bf{d}}^\beta)\q} \Lambda_{xx}^{\alpha \beta}(\q)\,.
\end{equation}
The positions ${\bf{d}}^\alpha$ of the center of bond $\alpha$ are given by:

\noindent
\begin{minipage}{0.47\textwidth}
	\begin{subequations}
		\begin{eqnarray}
		{\bf{d}}^1 &=& +0.5\hat{e}_x \\
		{\bf{d}}^2 &=& +0.5\hat{e}_x+0.5\hat{e}_y \\
		{\bf{d}}^3 &=& +0.5\hat{e}_x-0.5\hat{e}_y \\
		{\bf{d}}^4 &=& +0.5\hat{e}_x+\hat{e}_y \\
		{\bf{d}}^5 &=& +0.5\hat{e}_x+1.5\hat{e}_y \\ \nonumber
		\end{eqnarray}
	\end{subequations}
\end{minipage}
\hfill
\begin{minipage}{0.47\textwidth}
	\begin{subequations}
		\begin{eqnarray}
		{\bf{d}}^6 &=& +0.5\hat{e}_x+0.5\hat{e}_y \\
		{\bf{d}}^7 &=& +\hat{e}_x+\hat{e}_y \\
		{\bf{d}}^8 &=& +\hat{e}_x-\hat{e}_y \\
		{\bf{d}}^9 &=& +\hat{e}_x+2\hat{e}_x \\
		{\bf{d}}^{10} &=& +\hat{e}_x\, .  
		\end{eqnarray}
	\end{subequations}
\end{minipage}

\subsubsection{Numerical data of the current susceptibility}

\begin{figure}
	\begin{center}
		\includegraphics[width=0.6\columnwidth]{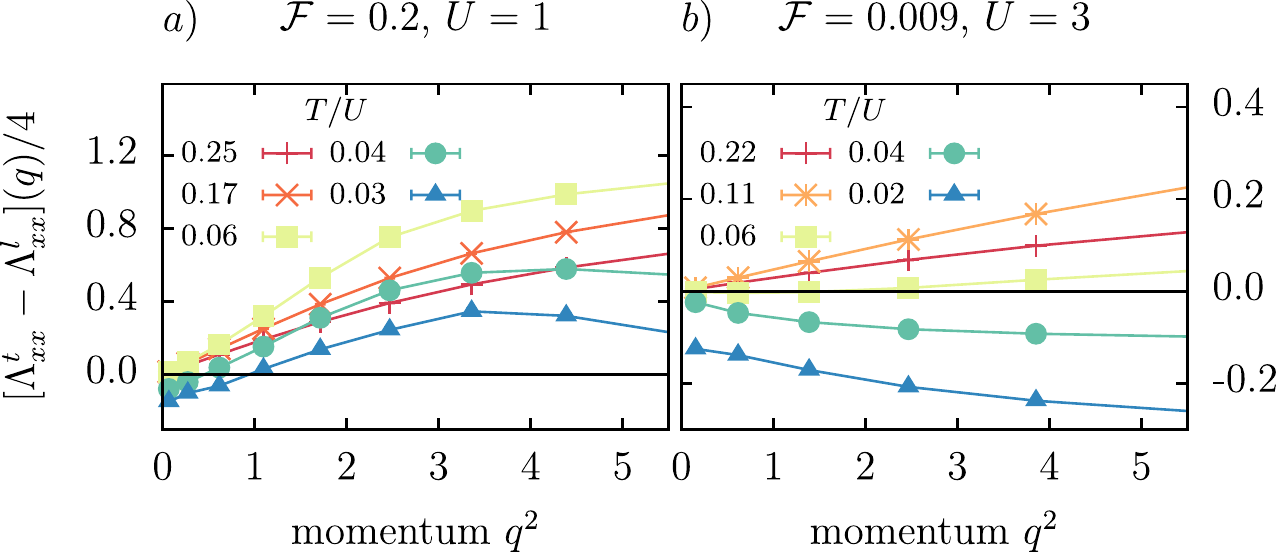}
	\end{center}
	\caption{Current-current correlation function $[\Lambda^t_{xx}-\Lambda^l_{xx}](q)/4$ as a function of momentum at various temperatures for the dispersive band (a) at coupling strength $U=1$ ($L=24$) and the flat band regime (b) with $U=3$ ($L=16$). 
	}
	\label{fig:Currents}
	
\end{figure}

The difference of the transverse and longitudinal contributions of the current susceptibility $[\Lambda^t_{xx}-\Lambda^l_{xx}](q)/4$ (recall the shorthand notation $\Lambda_{xx}^{\tn{t}}(q)=\Lambda_{xx}(q_x=0,q_y=q)$ and $\Lambda_{xx}^{\tn{l}}(q)=\Lambda_{xx}(q_x=q,q_y=0)$) is presented in Fig.~\ref{fig:Currents} at various temperatures for the dispersive (flat band) case on a lattice of linear size $L=24$ ($L=16$). At high temperatures, $1/4[\Lambda^t_{xx}-\Lambda^l_{xx}](q)$ is positive and increases monotonically with $q$ and extrapolates to zero for vanishing momentum $q$ as the system is not yet in the superconducting phase. The susceptibility first increases with decreasing temperatures, which is most noticeable at large momenta, reaches maximal values for $T/U\approx0.06$ ($T/U\approx0.11$) at temperatures slightly below $T_p/U=0.085$ ($T_p/U=0.145$) in the dispersive (flat-band) scenario, and then decreases with temperature. At low temperatures, the current susceptibility exhibits negative values and $\lim_{q\rightarrow 0}[\Lambda^t_{xx}-\Lambda^l_{xx}](q)/4 < 0$ when the system is in the superconducting phase.

Note the data for $T/U=0.04$ in the flat-band limit Fig.~\ref{fig:Currents}b. The correlation function still extrapolates to zero for vanishing momentum. Indeed, this temperature is above the critical temperature $T_c/U=0.025$ and hence in the normal conducting regime. However, all values are negative, $\frac{1}{4}[\Lambda^t_{xx}-\Lambda^l_{xx}](q) < 0$. Note that the magnetic orbital susceptibility is given by $\chi_{\tn{orb}}=\lim_{q\rightarrow 0} q^{-2} \big[\Lambda_{xx}^{\tn{t}}(q)-\Lambda_{xx}^{\tn{l}}(q)\big]$ and negative susceptibilities indicate a diamagnetic response.

\subsubsection{The phase stiffness}

We comment here on the definition of $D_s$ in Eq. (4) of the main text and the derivation of this equation (see, e.g., Ref.~\cite{SFcriteria}). The long-wavelength phase fluctuations of a two-dimensional superconductor are governed by an effective XY model, whose free energy is
\begin{equation}
F_s = \frac{1}{2} D_s \int d^2 r (\nabla \theta - 2e \mathbf{A})^2,
\end{equation}
where $D_s$ is the phase stiffness, $\theta$ is the phase of the superconducting order parameter, $e$ is the electron charge, and $\mathbf{A}$ is an external electromagnetic field. $D_s$ exhibits the well-known universal jump from $0$ to $2T_c/\pi$ at the BKT transition.

To compute $D_s$ from a microscopic model, we examine the response of the superconductor to an external magnetic field. In the superconducting phase, there are essentially no vortices in $\theta$, and we may choose a gauge where $\theta=\text{const}$. Then, the current density is given by
\begin{equation}
\mathbf{J} = -\frac{\partial f}{\partial \mathbf{A}} = -4e^2 D_s \mathbf{A}. 
\label{eq:A}
\end{equation}
Here, $\mathbf{A}$ should be interpreted as the transverse (divergence-free) part of the vector potential. On the other hand, we may compute the current $\mathbf{J}$ in response to a small external vector potential from the microscopic Hamiltonian, using linear response~\cite{SFcriteria}. Matching this computation to (\ref{eq:A}) yields Eq. (4) of the main text.

\subsection{Density of states near the Fermi level}

To probe for the opening of a gap in the single-particle spectrum, we show in Fig.~\ref{fig:PG_Tc}a the quantity $N_0=\int_\k \widetilde{G}(\k)$, where $\widetilde{G}(\k)$ is given by Eq. (5) of the main text, which acts as a proxy for the single-particle density of states near the Fermi level, as a function of temperature. Note that in the limit $T\rightarrow 0$, $N_0$ coincides wsith the single-particle density of states at the Fermi level.

$N_0(T)$ initially increases when the temperature is reduced, reaches a maximum at intermediate temperatures, and is then strongly suppressed at low temperatures, consistently with a fully gapped single particle spectrum. For the more dispersive band with $\mathcal{F}=0.2$, the maximum of $N_0(T)$ occurs at $T/U\approx0.06$, slightly above the critical temperature ($T_c/U=0.0445\pm0.0015$, see Fig.~\ref{fig:superconductivity}d). In contrast, for the narrower band with $\mathcal{F}=0.009$, the maximum in $N_0(T)$ occurs at $T/U\approx 0.08$, significantly above $T_c/U = 0.025\pm 0.001$. This indicates a pseudogap regime above $T_c$.

\begin{figure}
	\begin{center}
		\includegraphics[width=0.5\columnwidth]{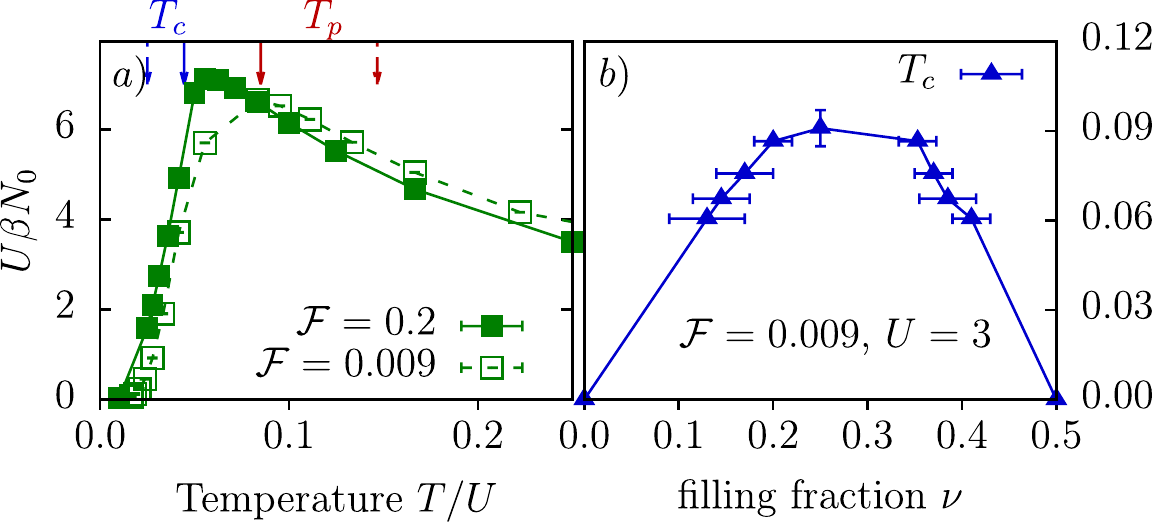}
	\end{center}
	\caption{
		Panel (a) depicts $N_0 = \int_k\tilde{G}(\k)$. 
		(b) $T_c$ as a function of the filling, $\nu$, for the flat band and $U=3$.
	}
	\label{fig:PG_Tc}
	
\end{figure}

\subsection{Dependence of $T_c$ on filling}
We have studied the dependence of $T_c$ on the filling, $\nu$, across the entire lower Chern band (Fig.~\ref{fig:PG_Tc}b). $T_c$ is found to have a broad maximum near half filling ($\nu=0.25$). This indicates that the physics we discuss here is not special to any particular value of the filling of the nearly-flat band. In the limit of $\nu\rightarrow0$ ($\nu\rightarrow0.5$), $\ie$ when the band is empty (full), we expect $T_c\rightarrow0$. 

\subsection{Approximate SU(2) symmetry}

The dramatic enhancement of the charge susceptibility with lowering temperature, especially in the case of a nearly-flat band (Fig. 3d of the main text), can be traced back to an approximate low-energy SU(2) symmetry that relates the superconducting and charge susceptibilities. The presence of this approximate symmetry in the low-energy effective Hamiltonian of flat band systems with local attractive interactions was pointed out in Ref.~\cite{Torma16b}. The symmetry becomes exact in the limit $U/\Delta_\tn{gap}\rightarrow0$, $T\rightarrow 0$, and a perfectly flat band, given that the single-particle projector onto the active band satisfies a certain condition, as explained below. For the sake of completeness, we provide a brief derivation of this result here, and discuss its consequences for our model.

{We start from an attractive Hubbard Hamiltonian with a single, perfectly flat partially-filled band. In the limit $U\ll \Delta_\tn{gap}$, the effective low-energy Hamiltonian to leading order in $U/\Delta_\tn{gap}$ can be obtained by projecting the interaction term
	\begin{equation}
	H_\tn{int}=-U\sum_{j}c_{j\uparrow}^{\dagger}c_{j\downarrow}^{\dagger}c_{j\downarrow}c_{j\uparrow}
	\end{equation}
	onto the flat band. The projected annihilation operator is given by
	\begin{equation}
	c_{j\sigma}=\sum_{\k}\psi_{\k\sigma}(\r_{j})c_{\k\sigma},
	\end{equation}
	where $\psi_{\k\sigma}(\r)$ is the Bloch wavefunction of the active band, and
	$\psi_{\k\downarrow}(\r)=\psi_{-\k\uparrow}^{*}(\r)$ by time reversal
	symmetry. We have used the fact that the Hamiltonian (1) conserves the $z$ component of the spin. 
	
	The projected interaction is thus of the form,
	\begin{equation}
	H_\tn{proj}=-U\sum_{j}\sum_{\k_{1}\dots \k_{4}}\psi_{\k_{1}\uparrow}^{*}(\r_{j})\psi_{\k_{2}\downarrow}^{*}(\r_{j})\psi_{\k_{3}\downarrow}(\r_{j})\psi_{\k_{4}\uparrow}(\r_{j})c_{\k_{1}\uparrow}^{\dagger}c_{\k_{2}\downarrow}^{\dagger}c_{\k_{3}\downarrow}c_{\k_{4}\uparrow}.
	\label{projham}
	\end{equation}
	Let us perform a particle-hole transformation in the active band as follows:
	\begin{align}
	c_{\k\uparrow} & \rightarrow d_{\k\uparrow}=c_{\k\uparrow},\nonumber \\
	c_{\k\downarrow} & \rightarrow d_{\k\downarrow}=c_{-\k\downarrow}^{\dagger},
	\end{align}
	under which the projected Hamiltonian in Eq.~\ref{projham} transforms to,
	\begin{align}
	H_\tn{proj}  \rightarrow H'_\tn{proj} & = U\sum_{j}\sum_{\k_{1}\dots \k_{4}}\psi_{\k_{1}\uparrow}^{*}(\r_{j})\psi_{\k_{2}\uparrow}(\r_{j})\psi_{\k_{3}\uparrow}^{*}(\r_{j})\psi_{\k_{4}\uparrow}(\r_{j})d_{\k_{1}\uparrow}^{\dagger}d_{\k_{3}\downarrow}^{\dagger}d_{\k_{2}\downarrow}d_{\k_{4}\uparrow}\nonumber \\
	& -U\sum_{j}\sum_{\k_{1},\k_{4}}\psi_{\k_{1}\uparrow}^{*}(\r_{j})\psi_{\k_{4}\uparrow}(\r_{j})\underset{P_\uparrow(\r_{j},\r_{j})}{\underbrace{\sum_{\k_{2}}\psi_{\k_{2}\uparrow}(\r_{j})\psi_{\k_{2}\uparrow}^{*}(\r_{j})}}d_{\k_{1}\uparrow}^{\dagger}d_{\k_{4}\uparrow}
	\end{align}
	where $P_\sigma(\r_{i},\r_{j}) = \sum_\k \psi_{\k,\sigma}(\r_i) \psi^*_{\k,\sigma}(\r_j)$ is a single-particle projection operator onto the active band with spin $\sigma$. If $P_\sigma(\r_{j},\r_{j})=\tn{const.}=P_{0}$, {\it i.e.} the diagonal of the projector onto the flat band is site-independent\footnote{This condition is termed the ``uniform pairing condition'' in~\cite{Torma16b}.}, we get 
	\begin{align}
	H'_\tn{proj} = U\sum_{j}\sum_{\k_{1}\dots \k_{4}}\psi_{\k_{1}\uparrow}^{*}(\r_{j})\psi_{\k_{2}\uparrow}(\r_{j})\psi_{\k_{3}\uparrow}^{*}(\r_{j})\psi_{\k_{4}\uparrow}(\r_{j})d_{\k_{1}\uparrow}^{\dagger}d_{\k_{3}\downarrow}^{\dagger}d_{\k_{2}\downarrow}d_{\k_{4}\uparrow} -P_0U\sum_{\k}d_{\k\uparrow}^{\dagger}d_{\k\uparrow}.
	\label{eq:Hprime0}
	\end{align}
	In our model (Eq. 1 in the main text), the condition $P_\sigma(\r_{j})=\tn{const.}$ is satisfied because the $A$ and $B$ sublattices of the square lattice are related by a inversion centered at a nearest-neighbor bond followed by a gauge transformation. 
	
	The second term in (\ref{eq:Hprime0}) is a conserved quantity, and does not affect the dynamics of the system. Working with a constant number of particles of either spin, it is a constant, and can be dropped. The first term can be written as
	\begin{equation}
	H'_\tn{proj}=U\sum_{j}\sum_{\sigma,\sigma'=\uparrow,\downarrow}\sum_{\k_{1}\dots \k_{4}}\psi_{\k_{1}\uparrow}^{*}(\r_{j})\psi_{\k_{3}\uparrow}^{*}(\r_{j})\psi_{\k_{2}\uparrow}(\r_{j})\psi_{\k_{4}\uparrow}(\r_{j})d_{\k_{1}\sigma}^{\dagger}d_{\k_{3}\sigma'}^{\dagger}d_{\k_{2}\sigma'}d_{\k_{4}\sigma},
	\label{eq:Hprime}
	\end{equation}
	where we note that the terms with $\sigma=\sigma'$ vanish because of the anti-symmetry with respect to $\k_{1}\longleftrightarrow \k_{3}$. The above form is manifestly SU(2) symmetric. In terms of the original electronic operators, the SU(2) generators are $I^z =   \frac{1}{2}\sum_{\k,\sigma} c^\dagger_{\k,\sigma} c_{\k,\sigma}$, $I^+  = \sum_{\k} c^\dagger_{\k,\uparrow} c^\dagger_{-\k,\downarrow}$, and $I^- = (I^+)^\dagger$. This shows that the pairing and charge susceptibilities of the effective Hamiltonian are equal. 
	
	This effective Hamitonian (\ref{eq:Hprime}) is positive semi-definite, and hence the fully polarized ``ferromagnetic'' state is an exact eigenstate. Thus there is a direct correspondence between the following observations, namely that the ground state for a half-filled flat band in a repulsive model (\ref{eq:Hprime}) is a completely polarized ferromagnetic state, while for an arbitrary filling of the flat band in the attractive model (\ref{projham}) it is the BCS state. The compressibility of the BCS ground state diverges~\cite{Torma16b}}.

\begin{figure}
	\begin{center}
		\includegraphics[width=0.7\columnwidth]{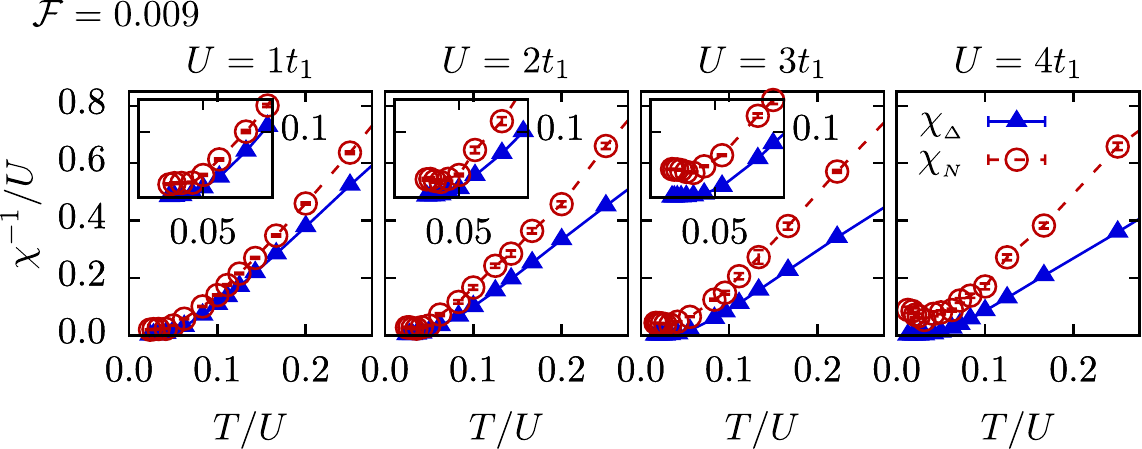}
	\end{center}
	\caption{Inverse charge and pairing susceptibilities for the flat band on $L=12$ lattices and various interaction strength~$U$.
	}
	\label{fig:EffSym}
	
\end{figure}

If the SU(2) symmetry were exact, the superconducting $T_c$ would vanish by the Mermin-Wagner theorem. There are different effects that break the SU(2) symmetry, and can favor either superconductivity or phase separation:  1. higher order corrections in $U/\Delta_\tn{gap}$ to the effective Hamiltonian; 2. a non-zero bandwidth $W$; 3. An extended interaction (beyond nearest-neighbor); and 3. non-zero temperature. The latter effect is exponentially small in $\Delta_\tn{gap}/T$, and is likely negligible near $T_c$ in our model. The long-wavelength fluctuations of the SU(2) order parameter can be described in terms of an effective non-linear Sigma model, whose free energy is written as
\begin{equation}
F_\tn{NLSM} = \frac{1}{2}\int d^{2}r \left[\rho_s (\nabla \vec{n})^2 + J_\perp (n_z)^2\right].
\label{eq:NLSM}
\end{equation}
Here, $\vec{n}$ is the three-component order parameter normalized such that $|\vec{n}|=1$, $\rho_s\propto U$ is the effective ``spin stiffness,'' and $J_\perp$ is a small anisotropy term, $|J_\perp| \sim \mathrm{max}(W,U^2/\Delta_\tn{gap},|V|)$, that describes the breaking of the SU(2) symmetry. 

In the absence of nearest-neighbor interactions, $V=0$, the superconducting $T_c$ in our model is non-zero, implying that the anisotropy is easy-plane, $J_\perp>0$. The superconducting transition temperature is then $T_c \propto \rho_s \log(\rho_s/J_\perp)$, implying that there should be a logarithmic correction to the relation $T_c\propto U$ for small $U$. Within our accuracy, we could not resolve such a correction, however (see Fig. 1c in the main text). Turning on small attractive nearest-neighbor interaction, $V=0.1U$, destroys the superconducting state and drives an instability towards phase separation (see Fig. 3b of the main text). In terms of the NLSM, this indicates that the addition of the nearest-neighbor attraction changes the sign of $J_\perp$ from positive to negative (easy-axis anisotropy).

In Fig.~\ref{fig:EffSym}, we present the charge and pairing susceptibilities for the flat band system on a $L=12$ lattice for various interaction strength $U=1-4$ as a function of temperature. At weak interactions {and {intermediate} temperatures}, the two susceptibilities are indeed almost identical and the charge susceptibility $\kappa$ is strongly enhanced at low temperatures. 
Increasing the coupling strength also increases both the deviations between the two susceptibilities and the finite value of $\chi_N^{-1}/U$ at low temperatures. This behavior is consistent with an emergent SU(2) symmetry in the limit $U/\Delta_\tn{gap}\rightarrow0$ and $T\rightarrow0$ as well as a small easy-plane symmetry breaking term that scales as $U^2/\Delta_\tn{gap}$.

\section{Competing orders}

Momentum resolved equal-time correlation functions are shown in Fig.~\ref{fig:CompOrder}. The pairing correlation function is defined in Eq.~\ref{CorrFct}, shown in the first column of Figs.~\ref{fig:CompOrder}a(b) for the dispersive (flat) band at high temperatures, slightly above $T_c$ and within the superconducting phase below $T_c$. At low temperatures, we recognize a sharp peak at ${\bf{q}}=0$ that signals the onset of $s$-wave singlet pairing order. In order to detect possible competing instabilities, we also study the charge ($N$), spin ($S^z$), and the bond-density ($N_b=1/4\sum_{\la i,j\ra}c^\dag_i c^{\phantom{\dag}}_j + \tn{h.c.}$) correlation functions. Note that the latter probes, e.g., for valence-bond-solid states. The absence of enhancement of these correlators at any non-zero wavevector indicates that there is no competing density wave ordering tendency. 
The enhancement of charge fluctuations and the associated tendency towards phase separation are indicated by the broad peak around $q=0$ in the density correlation function at intermediate temperatures in the flat band limit (Fig.~\ref{fig:CompOrder}b). However, as the temperature decreases, this feature disappears as the superconducting phase transition preempts this instability. For the band with a larger bandwidth, it is interesting to note that both in the density and bond-density response functions, there is a peak-like feature that develops at temperatures close to the superconducting $T_c$ (Fig.~\ref{fig:CompOrder}a) at wavevectors that roughly correspond to the `nesting' wavevectors of the original non-interacting Fermi surfaces. However, there is no diverging susceptibility in either of these two channels. The same response functions for the narrower band do not show these features, which is not surprising in the absence of the underlying Fermi surface (as deduced from the ``proxy" to the spectral function).

\end{widetext}
\end{document}